\documentclass[11pt,showpacs,preprint,preprintnumbers,nofootinbib,superscriptaddress,amsmath,amssymb]{revtex4}

\def\be{\begin{equation}}
\def\ee{\end{equation}}
\newcommand{\bea}{\begin{eqnarray}}
\newcommand{\eea}{\end{eqnarray}}

\usepackage{graphicx}% Include figure files
\usepackage{dcolumn}% Align table columns on decimal point
\usepackage{bm}% bold math
\usepackage{amssymb}
\usepackage{amsmath}
\usepackage{epsfig}    
\usepackage{color}
\usepackage{slashed}
\newcolumntype{I}{!{\vrule width 1.3pt}}
\begin{document} 
\title{Radiative Generation of the Lepton Mass}
\preprint{KIAS-P13063}

\author{Hiroshi Okada}
\email{hokada@kias.re.kr}
\affiliation{School of Physics, KIAS, Seoul 130-722, Korea}
\author{Kei Yagyu}
\email{keiyagyu@ncu.edu.tw}
\affiliation{Department of Physics and Center for Mathematics and Theoretical Physics,
National Central University, Chungli, Taiwan 32001, ROC}

\begin{abstract}

We propose a new mechanism where both Dirac masses for the charged-leptons and Majorana masses for neutrinos 
are generated via quantum levels. 
The charged-lepton masses are given by the vacuum expectation values (VEVs) of the Higgs doublet field and that of a triplet field. 
On the other hand, neutrino masses are generated by two VEVs of triplet Higgs fields. 
As a result, the hierarchy between the masses for charged-leptons and neutrinos 
can be explained by the triplet VEVs which have to be much smaller than the doublet VEV due to the constraint from the electroweak 
rho parameter. 
We construct a concrete model to realize this mechanism with discrete $\mathbb{Z}_2$ and $\mathbb{Z}_4$ symmetries, in which
masses for neutrinos and those for the muon and electron are generated at the one-loop level.  
As a bonus in our model, the deviation in the measured muon $g-2$ from the standard model prediction
can be explained by contributions of extra particle loops. 
Besides, the lightest $\mathbb{Z}_2$-odd neutral particle can be a dark matter candidate. 
The collider phenomenology is also discussed, especially focusing on doubly-charged scalar bosons which are necessary to 
introduce to occur our mechanism. 
\end{abstract}
\maketitle
%\newpage

\section{Introduction}

After the discovery of a Higgs boson at the CERN Large Hadron Collider (LHC)~\cite{Higgs}, 
we have completed to find all the elementary particles predicted in the Standard Model (SM), and 
the mechanism for the mass generation of the weak gauge bosons~\cite{Higgs_mech} have also been confirmed. 

However, there are puzzles in the mechanism for the generation of fermion masses in the SM.  
First, neutrino masses are exactly predicted to be zero, although
the establishment of existence of the neutrino oscillation
suggests that neutrinos have tiny but non-zero masses typically of order 0.1 eV. 
That is one of strong reasons that we need to consider physics beyond the SM.
Next, all the masses of charged fermions are given through the Yukawa interaction. 
Because only the vacuum expectation value (VEV) of the Higgs doublet field is the dimension full parameter in the SM, 
the magnitude of the measured fermion masses are accommodated by tuning the Yukawa coupling constants. 
That causes an unnatural situation especially for the charged-lepton sector. 
For instance, the electron mass is about 0.5 MeV which is obtained by taking the electron Yukawa coupling to be order of $10^{-5}$. 

Therefore, the questions are then following; (i) why lepton masses are quite small compared to the electroweak scale, i.e., $\mathcal{O}(100)$ GeV, 
(ii) why there is a large hierarchy between masses for neutrinos and those for charged-leptons.  
For instance, there is a difference of 6 orders between neutrino masses and the electron mass. 

One of the attractive scenarios to explain the smallness of neutrino masses are known as 
so-called radiative seesaw models~\cite{Radseesaw-1, Radseesaw-2, Radseesaw-3} in which neutrino masses are generated through quantum levels. 
In Ref.~\cite{1loop_fermion}, another type of the radiative seesaw model has been proposed, 
in which the masses for the first and second generation quarks and charged-leptons are generated at the one-loop level. 
However, the above scenarios cannot explain the smallness of the masses for both charged-leptons and neutrinos simultaneously, 
and those hierarchy.

In this paper, we propose a new radiative seesaw model. 
The Dirac masses for charged-leptons and Majorana masses for neutrinos 
are respectively generated from the one-loop induced dimension five operators as
\begin{align}
\frac{1}{16\pi^2}\frac{1}{\Lambda}\bar{L}_Le_R\Phi\Delta~~\text{and}~~\frac{1}{16\pi^2}\frac{1}{\Lambda'}\bar{L}_L^{c}L_L\Delta\Delta',
\label{eq1}
\end{align}
where $L_L$ ($e_R$) is the left (right) handed lepton field, and $\Phi$ is the isospin doublet scalar field. 
$\Delta$ and $\Delta'$ are respectively the isospin triplet scalar fields with the hypercharge $Y=0$ and $Y=1$.  
Typical mass scales are denoted by $\Lambda$ and $\Lambda'$. 
The mass hierarchy between charged-leptons and neutrinos can be explained 
by the VEVs of triplet scalar fields which have to be much smaller than the doublet VEV 
because of the constraint from the electroweak rho parameter.

This paper is organized as follows. 
In Sec.~II, we define our model, and we give the Lagrangian relevant to the generation of the lepton masses.
In Sec.~III, several observables in the lepton sector are calculated, e.g., masses for the charged-leptons and neutrinos, the
muon $g-2$, and lepton flavor violating (LFV) processes. 
Sec.~IV is devoted to disucss the collider phenomenology in our model. 
Conclusions and discussions are given in Sec.~V. 

\begin{center}
\begin{table}[t]
{\small
\hfill{}
\begin{tabular}{c|c|c|c|c|c||c|c|c|c|c|c}
\hline\hline Particle & $L_L^i=(L_L^e,L_L^\mu,L_L^\tau)$ & $ e_R^a=(e_R^{},\mu_R^{}) $ & $ \tau_R^{} $  &  $E_L^\alpha$ & $E_R^\alpha$ & $\Phi$ & $\xi$  & $\chi$  & $\eta$  & $\Phi_{3/2}$  & $S$ \\\hline
$SU(2)_I,~U(1)_Y$ & $\bm{2},-1/2$ & $\bm{1},-1$& $\bm{1},-1$ & $\bm{1},-1$ & $\bm{1},-1$ & $\bm{2},1/2$ & $\bm{3},0$  & $\bm{3},1$ & $\bm{2},1/2$  & $\bm{2},3/2$   & $\bm{1},0$ \\\hline
$U(1)_L$ & $-1$ & $-1$ & $-1$ & $-1$ & $-1$ & $0$ & $0$  & $0$ & $0$ & $0$ & $0$  \\\hline
$\mathbb{Z}_2$ & $+$ & $+$ & $+$ & $-$ & $-$ & $+$ & $+$  & $+$ & $-$ & $-$ & $-$  \\\hline
$\mathbb{Z}_4$ & $1$ & $-i$&$+i$& $1$ & $1$ & $-i$ & $-1$  & $-1$ & $1$ & $1$ & $-i$  \\\hline\hline
\end{tabular}
}
\hfill{}
\caption{The particle contents and their charge assignments under $SU(2)_I\times U(1)_Y\times U(1)_L\times \mathbb{Z}_2\times\mathbb{Z}_4$, where $U(1)_L$ is the global lepton number symmetry. 
The index $i$ ($a$) for $L_L$ ($e_R$) runs over the first, second and third (first and second) generation. }
\label{tab:1}
\end{table}
\end{center}

\section{The Model}
In order to realize this mechanism in the renormalizable theory, 
let us consider a following model whose particle contents and their charge assignments are shown in Table~\ref{tab:1}.
In our model, we introduce discrete $\mathbb{Z}_2$ (unbroken) and $\mathbb{Z}_4$ (softly-broken) symmetries in addition to the SM gauge symmetries
in order to forbid the tree level Dirac masses for charged-leptons and 
Majorana masses for neutrinos via the so-called type-II seesaw mechanism~\cite{HTM}. 
The charge of the global lepton number symmetry $U(1)_L$ is assigned to be $-1$ for all the lepton fields. 

We add two isospin $SU(2)_I$ singlet vector-like charged-leptons $E_L^\alpha$ and $E_R^\alpha$ with the $\alpha$ flavor. 
The scalar sector is also extended from the SM one, which is composed of 
two $SU(2)_I$ triplet scalar fields $\xi$ and $\chi$ with the hypercharge $Y=0$ and $Y=1$, respectively, 
and three doublet scalar fields $\Phi$ ($Y=1/2$), $\eta$ ($Y=1/2$) and $\Phi_{3/2}$ ($Y=3/2$) and 
a singlet neutral scalar field $S$. 
Among these scalar fields, $\Phi$, $\xi$ and $\chi$ are assigned to be $\mathbb{Z}_2$-even, so that they can have 
non-zero VEVs. 
Although we can construct a model in which all the charged-lepton masses are generated at the one-loop level, 
we here consider the case with radiative generation for the muon and electron masses. 
Thus, only the tauon mass is generated at the tree level.
That is the reason why the $\mathbb{Z}_4$ charge assignment only for the right-handed tauon field $\tau_R^{}$ 
is different from that for $e_R^a$ as seen in Table~\ref{tab:1}.

The Lagrangian relevant to the generation of the lepton masses is given by
\begin{align}
-\mathcal{L}&\supset M_{\alpha}\overline{E_R^\alpha} E_L^\alpha+y_\tau^i\overline{L_L^i}\Phi\tau_R^{}+y_S^{a\alpha}\overline{e_R^a}  E_L^\alpha S
+y_{\eta}^{i\alpha}\overline{L_L^i}\eta E_R^\alpha 
+y_{3/2}^{i\alpha} \overline{L_L^{ic}}(i\tau_2)\Phi_{3/2}E_L^\alpha +\rm{h.c.}\notag\\
&
+\kappa_e\eta^\dagger \xi \Phi S +\kappa_\nu \Phi_{3/2}^\dagger \xi \chi \eta +\rm{h.c.}, \label{Yuk}
\end{align}
where $M_{\alpha}$ is the mass of the $\alpha$-th vector-like lepton. 
We note that the $y^{i\alpha}_{3/2}$ term explicitly breaks the lepton number symmetry $U(1)_L$ as we can see 
the charge assignments shown in Table~\ref{tab:1}, 
which turns to be a source of Majorana masses for neutrinos. 
Thus, the Yukawa coupling $y^{i\alpha}_{3/2}$ could be expected to be tiny compared to 
the other Yukawa couplings given in Eq.~(\ref{Yuk}) due to the analogy of the canonical seesaw mechanism. 

The scalar fields can be parameterized as 
\begin{align}
&\Phi =\left[
\begin{array}{c}
\phi^+\\
\phi^0
\end{array}\right],\quad
\xi = \left[ 
\begin{array}{cc}
\frac{\xi^0}{\sqrt2} & \xi^+ \\
\xi^- & -\frac{\xi^0}{\sqrt2}
\end{array}\right],~
\chi = \left[ 
\begin{array}{cc}
\frac{\chi^+}{\sqrt2} & \chi^{++} \\
\chi^0 & -\frac{\chi^+}{\sqrt2}
\end{array}\right],\notag\\
&\Phi_{3/2} =\left[
\begin{array}{c}
\Phi_{3/2}^{++}\\
\Phi_{3/2}^+
\end{array}\right],~
\eta =\left[
\begin{array}{c}
\eta^+\\
\frac1{\sqrt2}(\eta_R+i\eta_I)
\end{array}\right],\quad S=\frac{1}{\sqrt{2}}(S_R+i S_I).  \label{component}
\end{align}
The VEVs for the above scalar fields are defined as 
$\langle \phi^0 \rangle =v_\phi/\sqrt{2}$, $\langle \chi^0 \rangle =v_\chi/\sqrt{2}$
and $\langle \xi^0 \rangle =v_\xi$, which are related to the Fermi constant $G_F$ by 
$v^2\equiv v_\phi^2+2v_\chi^2+4v_\xi^2=1/(\sqrt{2}G_F)=(246$ GeV$)^2$.
The electroweak rho parameter $\rho$ deviates from unity due to the non-zero value of $v_\xi$ and $v_\chi$ at the tree level as 
\begin{align}
\rho =\frac{v^2}{v^2+2v_\chi^2-4v_\xi^2}. 
\end{align}
The experimental value of the rho parameter is close to unity, 
so that both the triplet VEVs are constrained from the above to be of order 1 GeV\footnote{
If two triplet VEVs are aligned as $v_\chi=\sqrt{2}v_\xi$, then the deviation in the rho parameter from unity is cancelled, which 
has been known in the Georgi-Machacek model~\cite{GM}. 
In our model, we do not assume the alignment for the triplet VEVs to avoid such a fine tuning.}. 

Thanks to the unbroken $\mathbb{Z}_2$ parity, the scalar fields with $\mathbb{Z}_2$-even and those with $\mathbb{Z}_2$-odd do not mix with each other. 
The mass eigenstates for the CP-even, CP-odd and singly-charged scalar states with the $\mathbb{Z}_2$-odd parity 
can be obtained by introducing the three mixing angles as
\begin{align}
&\left(
\begin{array}{c}
S_R\\
\eta_R
\end{array}\right)=
R(\theta_R)
\left(
\begin{array}{c}
H_S\\
H_\eta
\end{array}\right)
,~~
\left(
\begin{array}{c}
S_I\\
\eta_I
\end{array}\right)=
R(\theta_I)
\left(
\begin{array}{c}
A_S\\
A_\eta
\end{array}\right),~~
\left(
\begin{array}{c}
\Phi_{3/2}^\pm\\
\eta^\pm
\end{array}\right)=
R(\theta_C)
\left(
\begin{array}{c}
H_{3/2}^\pm\\
H_\eta^\pm
\end{array}\right)
,\notag\\
&
\text{with}~~R(x)=\left(
\begin{array}{cc}
\cos x & -\sin x \\
\sin x & \cos x
\end{array}\right). 
\end{align}
The above mixing angles are expressed as 
\begin{align}
&\sin2\theta_R = -\frac{\kappa_ev_\phi v_\xi}{m^2_{H_\eta}-m^2_{H_S}},~~\sin2\theta_I=-\frac{\kappa_ev_\phi v_\xi}{m^2_{A_\eta}-m^2_{A_S}},
~~\sin2\theta_C = -\frac{\kappa_\nu v_\xi v_\chi}{m^2_{H_{\eta}^+}-m^2_{H_{3/2}^+} }, \label{mixing}
\end{align}
where $m_{\varphi}^2$ are the mass eigenvalues for the physical 
scalar bosons $\varphi$ ($=H_\eta$, $H_S$, $A_\eta$, $A_S$, $H_\eta^\pm$ and $H_{3/2}^\pm$)\footnote{The mass difference 
between a CP-even state and a CP-odd state with $\mathbb{Z}_2$-odd can be generated from the soft-breaking terms of $\mathbb{Z}_4$. 
For example, 
the mass difference between $S_R$ and $S_I$ can be explained by the $S^2$ term. 
}.
The lightest neutral scalar boson among $\varphi$ can be a dark matter candidate.  

We here comment on the SM-like Higgs boson which is mainly composed of Re$(\phi^0)$ in Eq.~(\ref{component}). 
Although there are mixings among Re$(\phi^0)$, $\xi^0$ and Re$(\chi^0)$, it can be neglected because of the suppression by the factor 
of $v_\xi/v$ or $v_\chi/v$. 
Thus, the current LHC results regarding the 126 GeV Higgs boson can be explained by the SM-like Higgs boson as well as by the SM Higgs boson.

\begin{figure}[t]
\begin{center}
 \includegraphics[width=140mm]{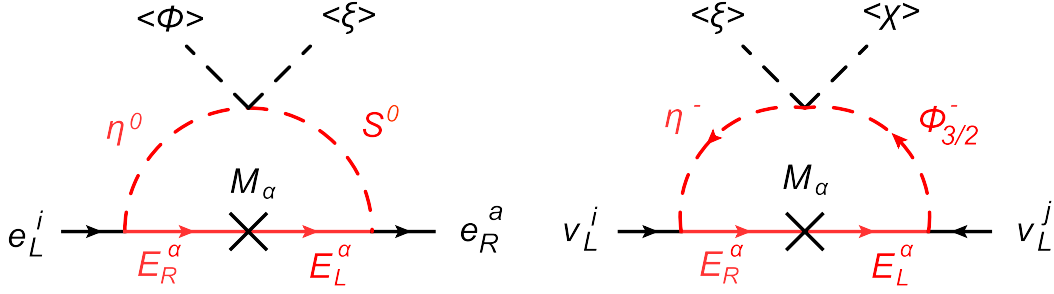}
   \caption{
The left and right panels respectively show the mass matrix for the charged-lepton and that 
for the neutrinos. The particles indicated by the red font have the $\mathbb{Z}_2$-odd parity.}
   \label{Rad_Lep}
\end{center}
\end{figure}

\section{Observables in the lepton sector}

After the spontaneous symmetry breaking, 
the mass matrices for the charged-leptons and neutrinos depicted in Fig.~\ref{Rad_Lep} are obtained by
\begin{align}
&M_{ia}^\ell=\sum_{\alpha}
\frac{M_{\alpha}}{64\pi^2}  y_\eta^{i\alpha*}y_S^{a\alpha}\sin2\theta_R F\left(\frac{m_{H_S}^2}{M_{\alpha}^2},\frac{m_{H_\eta}^2}{M_{\alpha}^2}\right) 
-[(H_S,H_\eta,\theta_R) \to (A_S,A_\eta,\theta_I)], \label{m_ell}\\
&M_{ij}^\nu=\sum_{\alpha}
\frac{M_{\alpha}}{32\pi^2}  (y_\eta^{i\alpha *} y_{3/2}^{j\alpha}+ y_\eta^{j\alpha *} y_{3/2}^{i\alpha}) 
\sin2\theta_C F\left(\frac{m_{H_{3/2}^+}^2}{M_{\alpha}^2},\frac{m_{H_\eta^+}^2}{M_{\alpha}^2}\right), \label{m_nu}
\end{align}
where
\begin{align}
F(x,y)=
\frac{-x\ln x+y\ln y +xy\ln\frac{x}{y}}{(1-x)(1-y)}.
\end{align}
From Eq.~(\ref{m_ell}), it is seen that 
all the matrix elements vanish when there is no mass difference between the mass of the CP-even scalar boson and 
that of the corresponding CP-odd scalar boson. 
In the following, we consider the case with $\alpha=3$. 
In fact, the case with $\alpha=2$ is enough to obtain two non-zero eigenvalues of $M^\ell$. 
However, in that case, there is no parameter sets to get the diagonal form of $M^\ell$ without introducing any unitary matrices. 
Non-zero values for the off diagonal elements in $M^\ell$
causes dangerous LFV processes such as $\mu\to e\gamma$, so that the case without such a off diagonal element is 
better to avoid the constraint from the $\mu\to e\gamma$ data.
We note that in general, there are $e$-$\tau$ and $\mu$-$\tau$ mixings at the tree level from $y_\tau$ couplings and at the one-loop level 
via the $y_\eta$ couplings.

To explicitly show how we can get the masses for the charged-leptons, we take the following assumptions
\footnote{
We can move to the basis with $y_\tau^1=y_\tau^2=0$ by the transformation of the left-handed lepton fields; i.e., $L_L\to O_L L_L'$, where $O_L$ is the orthogonal matrix. 
However, this transformation cannot be completely absorbed due to the different $Z_4$ charge assignment between $e_R$/$\mu_R$ and $\tau_R$.  
%Thus, this can slightly deviate the Yukawa matrix $y_S$ from the original one. 
We here simply take $y_\tau^1=y_\tau^2=0$ instead of the base transformation as in Eq.~(\ref{assump}). We would like to thank  referee to draw our attention to this issue.}
\begin{align}
&M_1=M_2=M_3=M,\notag\\
%&y_\eta^{21}= y_\eta^{12},\quad y_\eta^{32 }= y_\eta^{23},\quad 
%y_\eta^{31}= y_\eta^{13},\quad y_\eta^{33}= \frac{y_\eta^{13}y_\eta^{23}}{y_\eta^{12}},\notag\\
&y_\eta^{ij}=y_{\eta}^{33}=\bar{y}_\eta,\quad \text{for}~~i\neq j,\notag\\
%&y_S^{12}=y_S^{21}=0,\quad y_S^{13}=-\frac{y_{\eta}^{12}y_S^{11} }{y_\eta^{23}},\quad  
%y_S^{23}=-\frac{y_{\eta}^{12}y_S^{22} }{y_\eta^{13}}, \notag\\
&y_S^{12}=y_S^{21}=0,\quad y_S^{13}=-y_S^{11},\quad  y_S^{23}=-y_S^{22}, \notag\\
&y_\tau^1=y_\tau^2=0.  \label{assump}
\end{align}
Besides, all the elements in $y_\eta$ and $y_S$ are assumed to be real numbers. 
Under the above assumptions, the mass matrix for the charged-leptons can be expressed by 
\begin{align}
M^\ell =
\left[
%\begin{array}{ccc}
%\tilde{M}\left(y_{\eta}^{11}-\frac{y_{\eta}^{12}y_{\eta}^{13}}{y_{\eta}^{23}}\right)y_S^{11} & 0 & 0 \\
%0 & \tilde{M}\left(y_{\eta}^{22}-\frac{y_{\eta}^{12}y_{\eta}^{23}}{y_{\eta}^{13}}\right)y_S^{22} & 0 \\
%0 & 0 & \frac{v_\phi}{\sqrt{2}}y_\tau^3 \\
%\end{array}\right],
\begin{array}{ccc}
\tilde{M}\left(y_{\eta}^{11}-\bar{y}_\eta\right)y_S^{11} & 0 & 0 \\
0 & \tilde{M}\left(y_{\eta}^{22}-\bar{y}_\eta\right)y_S^{22} & 0 \\
0 & 0 & \frac{v_\phi}{\sqrt{2}}y_\tau^3 \\
\end{array}\right],
\end{align}
where 
\begin{align}
\tilde{M}\equiv \frac{M}{64\pi^2}
\sin2\theta_R F\left(\frac{m_{H_S}^2}{M^2},\frac{m_{H_\eta}^2}{M^2} \right)
-[(H_S,H_\eta,\theta_R) \to (A_S,A_\eta,\theta_I)]. 
\end{align}

\begin{figure}[t]
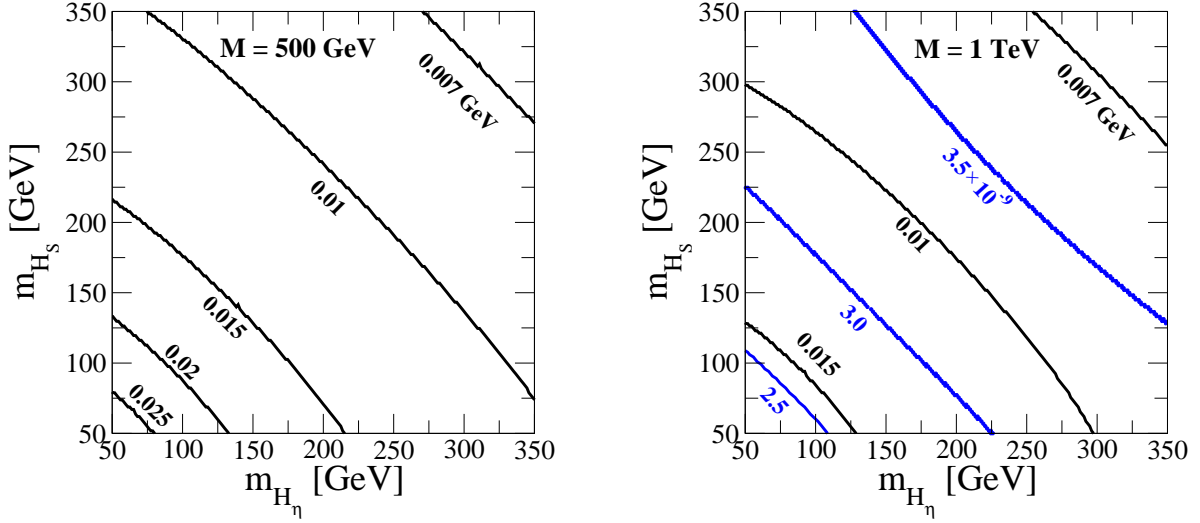

\begin{center}
\includegraphics[scale=0.4]{bm_500.eps}\hspace{10mm}
\includegraphics[scale=0.4]{bm_1000.eps}  
\caption{Contour plots for $\tilde{M}$ are shown by the black curves in the $m_{H_\eta}$-$m_{H_S}$ plane in the case of $\kappa_e v_\xi=10$ GeV. 
The left and right panels respectively show the cases with $M=500$ GeV and 1 TeV. 
In the right panel, the prediciton of $|\Delta a_\mu|$ using Eq.~(\ref{dela2}) is also shown as the blue contours.  }
\label{contour}
\end{center}
\end{figure}

In Fig.~\ref{contour}, we show the contour plots of $\tilde{M}$ in the $m_{H_\eta}$-$m_{H_S}$ plane 
in the case of $\kappa_e v_\xi=10$ GeV which 
contains in $\sin2\theta_R$ [see Eq.~(\ref{mixing})].
The mass of the vector-like lepton $M$ is taken to be 300 GeV (left panel), 500 GeV (center panel) and 1 TeV (right panel). 
We here show the contribution only from the CP-even scalar bosons; i.e., $H_\eta$ and $H_S$. 
The typical value of $\tilde{M}$ is seen to be of order $10^{-2}$ GeV in these cases, so that the Yukawa coupling constants
$(y_{\eta}^{22}-\bar{y}_\eta)y_S^{22}$ should be taken to be of order 10 to reproduce the muon mass about 0.1 GeV. 
We can obtain slightly larger values for $\tilde{M}$ in the cases with smaller $M$. 
The electron mass is simply obtained by taking the ratio $(y_{\eta}^{11}-\bar{y}_\eta)y_S^{11}$/$(y_{\eta}^{22}-\bar{y}_\eta)y_S^{22}$
to be $m_e/m_\mu\simeq 0.005$.

For the masses of neutrinos, there are additional suppression factors compared to the charged-lepton masses 
from $v_\chi/v $ and $y_{3/2}$ which is expected to be much smaller than $y_S$ and $y_\eta$
as we already mentioned before. 
%As can be seen in Eqs.~(\ref{mixing}), (\ref{m_ell}) and (\ref{m_nu}), 
%the neutrino masses are naturally tiny compared to the charged-lepton masses,
%because of the additional suppressions from $v_\chi/v \ll 1$ and $y_{3/2}^{i\alpha}\ll y_S^{a\alpha}, y_\eta^{i\alpha}$. 
Because there are enough degrees of freedom in $y_{3/2}$ to reproduce the neutrino mixings, 
we here give an order estimation for relevant parameters giving the order of neutrino masses; i.e., $\mathcal{O}(0.1)$ eV. 
When we take $M=1$ TeV, $m_\varphi=\mathcal{O}(100)$ GeV and $\kappa_\nu=\mathcal{O}(1)$, 
the product $v_\chi \times y_{3/2}$ should be taken to be $10^{-7}$ GeV.

\begin{figure}[t]
\begin{center}
 \includegraphics[width=120mm]{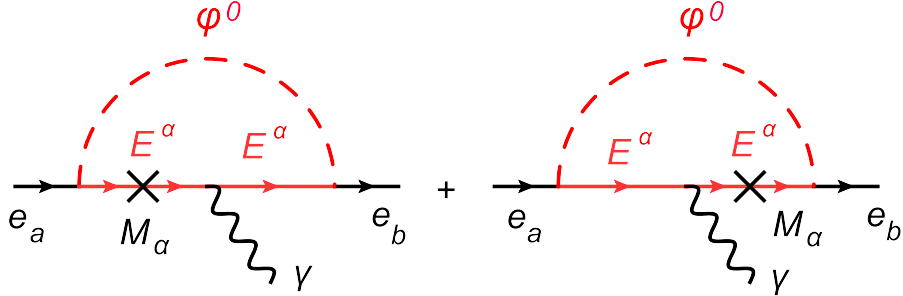}
   \caption{Dominant contributions to the $\ell_a\to \ell_b\gamma$ processes. 
$\varphi^0$ denote $H_\eta$, $H_S$, $A_\eta$ and $A_S$. }
   \label{gminus2}
\end{center}
\end{figure}

The muon anomalous magnetic moment has been 
measured at Brookhaven National Laboratory. 
The current average of the experimental results is given by~\cite{bennett}
\begin{align}
a^{\rm exp}_{\mu}=11 659 208.0(6.3)\times 10^{-10},\notag
\end{align}
which has a discrepancy from the SM prediction by $3.2\sigma$~\cite{discrepancy1} to $4.1\sigma$~\cite{discrepancy2} as
\begin{align}
\Delta a_{\mu}=a^{\rm exp}_{\mu}-a^{\rm SM}_{\mu}=(29.0 \pm
9.0\ {\rm to}\ 33.5 \pm
8.2)\times 10^{-10}. \notag
\end{align}
In our model, the vector-like charged-leptons and extra neutral scalar bosons  
can contribute to the $\ell_a\to \ell_b\gamma$ processes 
as shown in Fig.~\ref{gminus2}. 
The amplitude for these processes is calculated by
\begin{align}
&\Delta a_{ab}\simeq \sum_{\alpha=1}^3\frac{m_\mu \sin2\theta_R}{64\pi^2 M_{\alpha}}(y_S^{a\alpha*} y_\eta^{b \alpha}+y_S^{b\alpha*} y_\eta^{a \alpha})
\left[G\left(\frac{m_{H_S}^2}{M_{\alpha}^2}\right)-G\left(\frac{m_{H_\eta}^2}{M_{\alpha}^2}\right)\right]
-[(H_S,H_\eta,\theta_R) \to (A_S,A_\eta,\theta_I)], \label{dela}
\\
&\text{with}~~G(x)=\frac{1-4x +3 x^2-2 x^2 \ln x}{2(1-x)^3}, 
\end{align}
where terms proportional to $(y_S^{a\alpha})^2$ and $(y_{\eta}^{a\alpha})^2$ are neglected, because they 
give the contribution of order $m_\mu^2/M_{\alpha}^2$.
Moreover, we neglect the contribution from the $\Phi_{3/2}^{\pm\pm}$ loop diagram which is proportional to $(y_{3/2}^{a\alpha})^2$, 
and it is tiny enough, as mentioned in the above.
Under the assumptions given in Eq.~(\ref{assump}), 
Eq.~(\ref{dela}) can be rewritten by 
\begin{align}
&\Delta a_{ab}=2\left(\frac{m_\mu}{M^2}\right)\times \left(\frac{\tilde{G}}{\tilde{F}}\right)\times (M^\ell)_{ab}, \label{dela2}
\end{align}
where 
\begin{align}
\tilde{F}=F\left(\frac{m_{H_S}^2}{M^2},\frac{m_{H_\eta}^2}{M^2}\right),\quad 
\tilde{G}=G\left(\frac{m_{H_S}^2}{M^2}\right)-G\left(\frac{m_{H_\eta}^2}{M^2}\right). 
\end{align}
By taking $a=b=\mu$ in Eq.~(\ref{dela2}), we obtain the contribution to $\Delta a_{\mu}$ as
\begin{align}
\Delta a_{\mu}&= 2\left(\frac{m_\mu}{M}\right)^2\times \left(\frac{\tilde{G}}{\tilde{F}}\right), 
\end{align}
where $(M^\ell)_{\mu\mu}=m_\mu$ is used. 

In Fig.~\ref{contour} (right panel), 
we show the contour plots for the prediction of $|\Delta a_\mu|$ by the blue curves in the case of $M=1$ TeV and $\kappa_e\times v_\xi=10$ GeV. 
It can be seen that $|\Delta a_\mu|=3\times 10^{-9}$ can be explained when $m_{H_\eta}+m_{H_S}\simeq 275$ GeV which is corresponding to 
the case with $\tilde{M}\simeq 0.01$. 
Because $\Delta a_{ab}$ depends on $1/M^2$, smaller values of $M$ tend to get too large values of $\Delta a_{ab}$. 
For example, when we take $M=500$ GeV, the magnitude of $\Delta a_{ab}$ is about 5 times larger than that in the case with $M=1$ TeV.
Therefore, we can explain the discrepancy in the $g-2$ value in the case with $M=O(1)$ TeV in which we need to take 
$y_\eta^{22}\times y_S^{22}$ to be of order 10 to reproduce the muon mass as we discussed in the above.

%Because of the reproduction of the muon mass, 
%$\sin2\theta_{R(I)}$ and $M_{\alpha}$ are taken to be $\mathcal{O}$($10^{-2}$) and $\mathcal{O}$(1) TeV, respectively.
%Thus, when $y_S^{\mu\alpha}y_\eta^{\mu\alpha}[G(m_{H_S}^2/M_{\alpha}^2)-G(m_{H_\eta}^2/M_{\alpha}^2)]$ is order 0.1, we obtain $\Delta a_{\mu}={\cal O}$(10$^{-9}$). 

In the end of this section, we discuss the LFV process, especially for 
the $\mu\to e\gamma$, because it gives the most stringent constraint.  
The 95\% confidence level upper limit of the branching ratio is given by
${\cal B}\left(\mu\to e\gamma\right)\leq5.7\times 10^{-13}$ from the MEG experiment~\cite{meg}. 
In our model, the branching ratio is calculated as
 \begin{align}
 {\cal B} (\mu\to e\gamma)
&\simeq \frac{24\pi^3\alpha_{\text{em}}}{G_F^2m_\mu^4}|\Delta \alpha_{\mu e}|^2, \label{eq:meg}
 \end{align}
where $\alpha_{\mathrm{em}}$ is the fine structure constant.
The point is that the amplitude for $\mu\to e\gamma$ given in Eq.~(\ref{eq:meg}) depends on the same 
matrix $\Delta a$ as in the amplitude for the muon $g-2$.
Therefore, this constraint can be avoided as long as we choose the parameter sets which gives the diagonal form of $\Delta a_{ab}$. 
One example for such a parameter sets is given in Eq.~(\ref{assump}). 

\section{ Higgs phenomenology}

Finally, we give an outline of the collider phenomenology in our model.  
There are two pairs of doubly-charged scalar bosons $\chi^{\pm\pm}$ and $\Phi_{3/2}^{\pm\pm}$ 
which do not mix with each other due to the opposite assignment for the unbroken $\mathbb{Z}_2$ charge. 
In addition, the isospin charge is also different; namely, 
$\chi^{\pm\pm}$ and $\Phi_{3/2}^{\pm\pm}$ respectively come from the triplet field and the doublet field, 
so that their properties can be quite different.  
The phenomenology of $\chi^{\pm\pm}$ is similar\footnote{
In our model, $\chi$ does not couple to the lepton fields, so that 
$\chi^{\pm\pm}$ cannot decay into the same-sign dilepton. 
Such a same-sign lepton decay of $\chi^{\pm\pm}$ can be allowed in the minimal Higgs triplet model when the triplet VEV is taken to be smaller than 
about 0.1 MeV.} 
to that in the minimal Higgs triplet model~\cite{HTM}.  
When the mass of $\chi^{\pm\pm}$ is the smallest among the component scalar fields in $\chi$, $\chi^{\pm\pm}$ can mainly decay into the same sign diboson $\chi^{\pm\pm}\to W^\pm W^\pm$. 
Collider signatures in such a scenario have been discussed in Refs.~\cite{diboson}. 
Recently in Ref.~\cite{KYY}, the lower mass bound for $\chi^{\pm\pm}$ has been found to be about 60 GeV by using the current LHC data. 

On the other hand, the decay of $\Phi_{3/2}^{\pm\pm}$ depends on the mass spectrum of the $\mathbb{Z}_2$-odd particles\footnote{
The collider phenomenology for $\Phi_{3/2}^{\pm\pm}$ has been discussed in Ref.~\cite{3over2}, but its decay process is different from 
that of $\Phi_{3/2}^{\pm\pm}$ in our model, because of the particle content. }. 
When the mass of $E_\alpha$ is smaller than that of $\Phi_{3/2}^{\pm\pm}$, 
the decay process $\Phi_{3/2}^{\pm\pm}\to \ell^\pm E_\alpha^\pm \to \ell^\pm  \mu^\pm \varphi^0$ is allowed, where  
$\varphi^0$ is a neutral $\mathbb{Z}_2$-odd scalar boson, and $\ell^\pm$ represent $e^\pm$, $\mu^\pm$ or $\tau^\pm$. 
In this process, we note that the decay of $E_\alpha$ is determined by the structure of the charged-lepton mass matrix, so that 
$E_\alpha$ prefer to decay into a muon and $\varphi^0$. 
Thus, at least one of two leptons with the same-sign is muon in the signal event from the decay of $\Phi_{3/2}^{\pm\pm}$. 
When the mass of $E_\alpha$ is larger than that of $\Phi_{3/2}^{\pm\pm}$, which is rather preferred scenario from the generation of 
lepton masses, 
the decay process can be $\Phi_{3/2}^{\pm\pm}\to \chi^{\pm\pm}\varphi^0$ or $\Phi_{3/2}^{\pm\pm}\to \chi^{\pm}\varphi^0W^{+*}$ via 
the $\kappa_\nu$ coupling defined in Eq.~(\ref{Yuk}). 
The decay of $\chi^{\pm\pm}$ are already discussed in the above, and that of $\chi^{\pm}$ depend on various parameters such as 
the mixing between $\xi^{\pm}$. 
When the mixing is not so important, $\chi^{\pm}$ can decay into $W^\pm Z$. 

\section{Conclusions and Discussions }%

%\begin{center}
%\begin{table}[t]
%{\small
%\hfill{}
%\begin{tabular}{c||c|c|c|c|c|c|c|c|c|c}\hline\hline 
% & \multicolumn{2}{c|}{Masses}   & \multicolumn{2}{c|}{Triplet VEVs}   & 
%\multicolumn{6}{c}{Couplings constants}       \\\hline
%Parameters & $M$ & $m_\varphi$  & $v_\xi$  & $v_\chi$ & $y_\eta^{22}y_S^{22}$ & $y_\eta^{11}y_S^{11}$ & $\bar{y}_\eta$ & $y_{3/2}$&
%$\kappa_e$ & $\kappa_\nu$   \\\hline
%Typical values & $\simeq$ 1 TeV & $\mathcal{O}(100)$ GeV & $\simeq 3$ GeV & $\mathcal{O}(1)$ GeV & $\simeq 10$ & $\simeq 0.05$ & $\simeq 0$ & 
%$\mathcal{O}(10^{-7})$&$\simeq 3$ & $\mathcal{O}(1)$   \\\hline\hline
%
%\end{tabular}
%}
%\hfill{}
%\caption{Typical values of the parameters giving correct orders of the observables in the lepton sector under the assumptions 
%in Eq.~(\ref{assump}). }
%\label{tab:2}
%\end{table}
%\end{center}

\begin{center}
\begin{table}[t]
{\small
\hfill{}
\begin{tabular}{c||c|c|c|c|c|c|c}\hline\hline 
Parameters & $M$ & $m_\varphi$  & $v_\xi\times \kappa_e$  & $v_\chi\times y_{3/2}$ & $y_\eta^{22}y_S^{22}$ & $y_\eta^{11}y_S^{11}$ & $\kappa_\nu$   \\\hline
Typical values & $\simeq$ 1 TeV & $\mathcal{O}(100)$ GeV & $\simeq 10$ GeV & $\mathcal{O}(10^{-7})$ GeV & $\simeq 10$ & $\simeq 0.05$ & $\mathcal{O}(1)$   \\\hline\hline
\end{tabular}
}
\hfill{}
\caption{Typical values of the parameters giving correct orders of the observables in the lepton sector under the assumptions 
in Eq.~(\ref{assump}). }
\label{tab:2}
\end{table}
\end{center}

We have constructed the new type of the radiative seesaw model which provides masses for charged-leptons and neutrinos
at the one-loop level in a TeV scale physics. 
In our model, the mass hierarchy between charged-leptons and neutrinos 
is naturally explained due to the smallness of the VEVs of triplet Higgs fields 
and the lepton number violating Yukawa coupling constant. 
In Table~\ref{tab:2}, we summarize 
the typical values of the parameters giving correct orders of the masses of charged-leptons and neutrinos
under the assumptions in Eq.~(\ref{assump}). 
We have found that the deviation in the measured muon $g-2$ from the SM prediction 
can be explained in the parameter sets given in Table~\ref{tab:2} 
by the loop effects of the vector-like leptons $E_\alpha$ and extra neutral $\mathbb{Z}_2$-odd scalar bosons. 
The decay rate of $\mu\to e\gamma$ can be neglected in this case due to the diagonal structure of the $\Delta a$ matrix given in Eq.~(\ref{dela2}). 
We then have discussed the collider phenomenology focusing on two pairs of doubly-charged scalar bosons. 

Our model has several bosonic dark matter candidates; namely, the lightest neutral $\mathbb{Z}_2$-odd scalar boson $H_\eta$, $A_\eta$, 
$H_S$ or $A_S$.
In the case of $SU(2)_I$ doublet-like scalar boson ($H_\eta$ or $A_\eta$), 
it has been already discussed in Ref.~\cite{DM}, 
in which the dark matter mass is in favor of at around the half of the observed Higgs boson mass, i.e., about 63 GeV, from the constraint of WMAP and the direct detection search in XENON100. 
%%%
In the case  of $SU(2)_I$ singlet-like boson ($H_S$ or $A_S$), it could explain the Fermi-LAT 130 GeV $\gamma$-ray excess~\cite{fermilat130} based on the model in Ref.~\cite{toma} if the mixing angles $\theta_R$ and $\theta_I$ are tiny enough to satisfy the current upper bound reported by XENON LUX~\cite{xenonlux}.  
Such a situation can be achieved in our framework.
%
%In our model, although only the tauon mass is obtained at the tree level,  we can extend our model so as to generate the tauon mass at the one-loop %level as well as the muon and electron masses. However, in that version of the model, it is difficult to explain the anomaly of $\Delta a_\mu$.
%
\\\\
\noindent
$Acknowledgments$

The authors would like to thank Yuji Kajiyama for fruitful discussions.
H.O. thanks to Prof. Eung-Jin Chun and Dr. Takashi Toma for useful discussions.
K.Y. was supported in part by the National Science Council of R.O.C. under Grant No. NSC-101-2811-M-008-014.
%%%%%%%%%%%%%%%%%%%%%%%%%%%%%%%%%%%%%%%%%%%%%%%%%%%%%%%%%%%%%%%%%%%%%%%%%%%%%%%%

%%%%%%%%%%%%%%%%%%%%%%%%%%%%%%%%%%%%%
%%%%%%%%  references  %%%%%%%%%%%%%%%
%%%%%%%%%%%%%%%%%%%%%%%%%%%%%%%%%%%%%


\begin{thebibliography}{99}

\bibitem{Higgs}
  G.~Aad {\it et al.}  [ATLAS Collaboration],
  %``Observation of a new particle in the search for the Standard Model Higgs boson with the ATLAS detector at the LHC,''
  Phys.\ Lett.\ B {\bf 716}, 1 (2012);
  S.~Chatrchyan {\it et al.}  [CMS Collaboration],
  %``Observation of a new boson at a mass of 125 GeV with the CMS experiment at the LHC,''
  Phys.\ Lett.\ B {\bf 716}, 30 (2012).
  %%CITATION = ARXIV:1207.7235;%%

%\cite{Higgs:1964pj}
\bibitem{Higgs_mech} 
  P.~W.~Higgs,
  %``Broken Symmetries and the Masses of Gauge Bosons,''
  Phys.\ Rev.\ Lett.\  {\bf 13}, 508 (1964);
  %%CITATION = PRLTA,13,508;%%
  %2448 citations counted in INSPIRE as of 14 Nov 2013
%\cite{Higgs:1966ev}
  P.~W.~Higgs,
  %``Spontaneous Symmetry Breakdown without Massless Bosons,''
  Phys.\ Rev.\  {\bf 145}, 1156 (1966);
  %%CITATION = PHRVA,145,1156;%%
  %1902 citations counted in INSPIRE as of 14 Nov 2013
%\cite{Englert:1964et}
  F.~Englert and R.~Brout,
  %``Broken Symmetry and the Mass of Gauge Vector Mesons,''
  Phys.\ Rev.\ Lett.\  {\bf 13}, 321 (1964);
  %%CITATION = PRLTA,13,321;%%
  %2232 citations counted in INSPIRE as of 14 Nov 2013
%\cite{Guralnik:1964eu}
  G.~S.~Guralnik, C.~R.~Hagen and T.~W.~B.~Kibble,
  %``Global Conservation Laws and Massless Particles,''
  Phys.\ Rev.\ Lett.\  {\bf 13}, 585 (1964).
  %%CITATION = PRLTA,13,585;%%
  %1775 citations counted in INSPIRE as of 14 Nov 2013
\bibitem{Radseesaw-1}
As for one-loop neutrino models: see, {\it e.g.}, 
 A.~Zee,
 %``A Theory Of Lepton Number Violation, Neutrino Majorana Mass, And
 %Oscillation,''
 Phys.\ Lett.\  B {\bf 93}, 389 (1980) 
 [Erratum-ibid.\  B {\bf 95}, 461  (1980) ];
   E.~Ma,
  %``Verifiable radiative seesaw mechanism of neutrino mass and dark matter,''
  Phys.\ Rev.\  D {\bf 73}, 077301  (2006).
 %%% Two loop level
 \bibitem{Radseesaw-2}
As for two-loop neutrino models: see, {\it e.g.}, 
 A.~Zee,
 %``Charged Scalar Field And Quantum Number Violations,''
 Phys.\ Lett.\  B {\bf 161}, 141 (1985);
 %%%
 A.~Zee,
 %``Quantum Numbers Of Majorana Neutrino Masses,''
 Nucl.\ Phys.\ B {\bf 264}, 99  (1986); 
 %%%
 K.~S.~Babu,
 %``MODEL OF 'CALCULABLE' MAJORANA NEUTRINO MASSES,''
 Phys.\ Lett.\ B {\bf 203}, 132  (1988); 
   M.~Aoki, J.~Kubo and H.~Takano,
  %``Two-loop radiative seesaw with multicomponent dark matter explaining the possible gamma excess in the Higgs boson decay and at the Fermi LAT,''
  Phys.\ Rev.\ D {\bf 87}, 116001 (2013); 
    Y.~Kajiyama, H.~Okada and K.~Yagyu,
  %``Two Loop Radiative Seesaw Model with Inert Triplet Scalar Field,''
  Nucl.\ Phys.\ B {\bf 874}, 198 (2013).
 %%% three-loop %%%%
 \bibitem{Radseesaw-3}
 As for three-loop neutrino models: see, {\it e.g.}, 
  L.~M.~Krauss, S.~Nasri and M.~Trodden,
  %``A model for neutrino masses and dark matter,''
  Phys.\ Rev.\  D {\bf 67}, 085002  (2003);
  %%%
 M.~Aoki, S.~Kanemura and O.~Seto,
  Phys.\ Rev.\ Lett.\  {\bf 102}, 051805 (2009);
  M.~Gustafsson, J.~M.~No and M.~A.~Rivera,
  Phys.\ Rev.\ Lett.\  {\bf 110}, 211802 (2013);
  Y.~Kajiyama, H.~Okada and K.~Yagyu,  JHEP {\bf10}, 196 (2013). 

\bibitem{1loop_fermion} 
  E.~Ma, D.~Ng, J.~T.~Pantaleone and G.~-G.~Wong,
  %``One Loop Induced Fermion Masses and Exotic Interactions in a Standard Model Context,''
  Phys.\ Rev.\ D {\bf 40}, 1586 (1989);
  %%CITATION = PHRVA,D40,1586;%%
  %7 citations counted in INSPIRE as of 05 Feb 2014
%\cite{Ma:1989cn}
  E.~Ma, D.~Ng and G.~-G.~Wong,
  %``Muon - Electron Transitions In Models Of Radiative Lepton Masses,''
  Z.\ Phys.\ C {\bf 47}, 431 (1990).
  %%CITATION = ZEPYA,C47,431;%%
  %4 citations counted in INSPIRE as of 05 Feb 2014 


\bibitem{HTM}
%\cite{Cheng:1980qt}
%\bibitem{Cheng:1980qt}
 T.~P.~Cheng and L.~F.~Li,
 %``Neutrino Masses, Mixings And Oscillations In SU(2) X U(1) Models Of
 %Electroweak Interactions,''
 Phys.\ Rev.\  D {\bf 22}, 2860 (1980);
 J.~Schechter and J.~W.~F.~Valle,
 %``Neutrino Masses In SU(2) X U(1) Theories,''
 Phys.\ Rev.\  D {\bf 22}, 2227 (1980);
 %%CITATION = PHRVA,D22,2227;%%
%\cite{Lazarides:1980nt}
%\bibitem{Lazarides:1980nt}
  G.~Lazarides, Q.~Shafi and C.~Wetterich,
  %``Proton Lifetime and Fermion Masses in an SO(10) Model,''
  Nucl.\ Phys.\  B {\bf 181}, 287 (1981);
  %%CITATION = NUPHA,B181,287;%%
  %\cite{Mohapatra:1980yp}
%\bibitem{Mohapatra:1980yp}
  R.~N.~Mohapatra and G.~Senjanovic,
  %``Neutrino Masses and Mixings in Gauge Models with Spontaneous Parity
  %Violation,''
  Phys.\ Rev.\  D {\bf 23}, 165 (1981);
  %%CITATION = PHRVA,D23,165;%%
  %\cite{Magg:1980ut}
%\bibitem{Magg:1980ut}
  M.~Magg and C.~Wetterich,
  %``NEUTRINO MASS PROBLEM AND GAUGE HIERARCHY,''
  Phys.\ Lett.\  B {\bf 94}, 61 (1980).
  %%CITATION = PHLTA,B94,61;%%

\bibitem{GM}
  H.~Georgi and M.~Machacek,
  %``Doubly Charged Higgs Bosons,''
  Nucl.\ Phys.\ B {\bf 262}, 463 (1985);
  M.~S.~Chanowitz and M.~Golden,
  %``HIGGS BOSON TRIPLETS WITH M (W) = M (Z) cos theta omega,''
  Phys.\ Lett.\ B {\bf 165}, 105 (1985).
  %%CITATION = PHLTA,B165,105;%%

\bibitem{bennett}
G. W. Bennett et al, [Muon $g-2$ Collaboration], Phys. Rev. D {\bf 73}, 072003 (2006).

\bibitem{discrepancy1}
%\cite{Jegerlehner:2009ry}\bibitem{Jegerlehner:2009ry} 
  F.~Jegerlehner and A.~Nyffeler,
  %``The Muon g-2,''
  Phys.\ Rept.\  {\bf 477}, 1 (2009).
%  [arXiv:0902.3360 [hep-ph]].
  %%CITATION = ARXIV:0902.3360;%%
  %296 citations counted in INSPIRE as of 09 Nov 2013
  
%\cite{Benayoun:2011mm}
\bibitem{discrepancy2} 
  M.~Benayoun, P.~David, L.~Delbuono and F.~Jegerlehner,
  %``Upgraded Breaking Of The HLS Model: A Full Solution to the $\tau^-e^+e^-$ and $\phi$ Decay Issues And Its Consequences On g-2 VMD Estimates,''
  Eur.\ Phys.\ J.\ C {\bf 72}, 1848 (2012). 
%  [arXiv:1106.1315 [hep-ph]].
  %%CITATION = ARXIV:1106.1315;%%  

%\cite{Adam:2013mnn}
\bibitem{meg} 
%\cite{Adam:2013mnn}\bibitem{Adam:2013mnn} 
  J.~Adam {\it et al.}  [MEG Collaboration],
  %``New constraint on the existence of the mu+-> e+ gamma decay,''
  Phys.\ Rev.\ Lett.\  {\bf 110}, 201801 (2013).
%   [arXiv:1303.0754 [hep-ex]].
  %%CITATION = ARXIV:1303.0754;%%
  %58 citations counted in INSPIRE as of 14 Nov 2013


\bibitem{diboson}
%  K.~Huitu, J.~Maalampi, A.~Pietila and M.~Raidal,
  %``Doubly charged Higgs at LHC,''
%  Nucl.\ Phys.\ B {\bf 487}, 27 (1997);
%  [hep-ph/9606311].
  T.~Han, B.~Mukhopadhyaya, Z.~Si and K.~Wang,
  %``Pair production of doubly-charged scalars: Neutrino mass constraints and signals at the LHC,''
  Phys.\ Rev.\ D {\bf 76}, 075013 (2007);
%  [arXiv:0706.0441 [hep-ph]];
  P.~Fileviez Perez, T.~Han, G.~-y.~Huang, T.~Li and K.~Wang,
  %``Neutrino Masses and the CERN LHC: Testing Type II Seesaw,''
  Phys.\ Rev.\ D {\bf 78}, 015018 (2008).
  C.~-W.~Chiang, T.~Nomura and K.~Tsumura,
  %``Search for doubly charged Higgs bosons using the same-sign diboson mode at the LHC,''
  Phys.\ Rev.\ D {\bf 85}, 095023 (2012). 
%  [arXiv:1202.2014 [hep-ph]].

\bibitem{KYY} 
  S.~Kanemura, K.~Yagyu and H.~Yokoya,
  %``First constraint on the mass of doubly-charged Higgs bosons in the same-sign diboson decay scenario at the LHC,''
  Phys.\ Lett.\ B {\bf 726}, 316 (2013). 
%  [arXiv:1305.2383 [hep-ph]].

%\cite{Aoki:2011yk}
\bibitem{3over2} 
  M.~Aoki, S.~Kanemura and K.~Yagyu,
  %``Doubly-charged scalar bosons from the doublet,''
  Phys.\ Lett.\ B {\bf 702}, 355 (2011)
  [Erratum-ibid.\ B {\bf 706}, 495 (2012)].
%  [arXiv:1105.2075 [hep-ph]].
  %%CITATION = ARXIV:1105.2075;%%

%\bibitem{PDG}
%Beringer et al. (Particle Data Group),~Phys.\ Rev.\ D~{\bf 86}, 010001 (2012).



\bibitem{DM}
%\cite{Kajiyama:2012xg}\bibitem{Kajiyama:2012xg} 
  Y.~Kajiyama, H.~Okada and T.~Toma,
  %``Light Dark Matter Candidate in B-L Gauged Radiative Inverse Seesaw,''
  Eur.\ Phys.\ J.\ C {\bf 73}, 2381 (2013).
%  [arXiv:1210.2305 [hep-ph]].
  %%CITATION = ARXIV:1210.2305;%%
  %7 citations counted in INSPIRE as of 10 Nov 2013

\bibitem{fermilat130} 
 %\cite{Bringmann:2012vr}\bibitem{Bringmann:2012vr} 
  T.~Bringmann, X.~Huang, A.~Ibarra, S.~Vogl and C.~Weniger,
  %``Fermi LAT Search for Internal Bremsstrahlung Signatures from Dark Matter Annihilation,''
  JCAP {\bf 1207}, 054 (2012);
%\cite{Weniger:2012tx}\bibitem{Weniger:2012tx} 
  C.~Weniger,
  %``A Tentative Gamma-Ray Line from Dark Matter Annihilation at the Fermi Large Area Telescope,''
  JCAP {\bf 1208}, 007 (2012).
%  [arXiv:1204.2797 [hep-ph]].
  %%CITATION = ARXIV:1204.2797;%%
  %219 citations counted in INSPIRE as of 14 Nov 2013


%\cite{Toma:2013bka}
\bibitem{toma} 
  T.~Toma,
  %``Internal Bremsstrahlung Signature of Real Scalar Dark Matter and Consistency with Thermal Relic Density,''
  Phys.\ Rev.\ Lett.\  {\bf 111}, 091301 (2013);
%  [arXiv:1307.6181 [hep-ph]].
  %%CITATION = ARXIV:1307.6181;%%
  %5 citations counted in INSPIRE as of 10 Nov 2013
  %\cite{Giacchino:2013bta}\bibitem{Giacchino:2013bta} 
  F.~Giacchino, L.~Lopez-Honorez and M.~H.~G.~Tytgat,
  %``Scalar Dark Matter Models with Significant Internal Bremsstrahlung,''
JCAP {\bf 1310}, 025 (2013).
%  arXiv:1307.6480 [hep-ph].
  %%CITATION = ARXIV:1307.6480;%%
  %6 citations counted in INSPIRE as of 24 Nov 2013
  
  
%\cite{Aprile:2012nq}
\bibitem{xenonlux} 
D.~S.~Akerib {\it et al.}  [LUX Collaboration],
  %``First results from the LUX dark matter experiment at the Sanford Underground Research Facility,''
  arXiv:1310.8214 [astro-ph.CO].
  %%CITATION = ARXIV:1310.8214;%%




\end{thebibliography}
\end{document}